\documentclass[
aps,
pra,
reprint,
superscriptaddress
]{revtex4-2}

\usepackage{graphicx}  % Include figure files
\usepackage{subfigure}
\usepackage{multirow}
\usepackage{xcolor} % color the text
\usepackage{upgreek}
\usepackage{comment}

\usepackage{fancyhdr}
\usepackage{longtable}
\usepackage[T1]{fontenc}
\usepackage{dcolumn}   % Align table columns on decimal point

\usepackage{bm}        % bold math
\usepackage{amsfonts}  % Common math fonts
\usepackage{amsmath}   % Common math functions
\usepackage{amssymb}   % Common math symbols
\begin{document}

\title{Quantum memory in an effective $\Lambda$-type system using dynamic detuning control of a superconducting qubit pair}

\author{Hsin Chang}
\affiliation{Department of Physics, National Central University, Taoyuan 320317, Taiwan}
\affiliation{Department of Electrical Engineering and Computer Science, Massachusetts Institute of Technology, Cambridge, MA 02139, USA}

\author{Kai-I Chu}
\email[Correspondence to: ]{kaiichu0903@gmail.com}
\affiliation{Department of Applied Physics, Graduate School of Engineering, The University of Tokyo, Bunkyo-ku, Tokyo 113-8656, Japan}

\author{Ite A. Yu}
\affiliation{Department of Physics, National Tsing Hua University, Hsinchu 300044, Taiwan}
\affiliation{Center for Quantum Science and Technology, National Tsing Hua University, Hsinchu 300044, Taiwan}
\affiliation{National Center for Excellence in Quantum Information Science and Engineering, National Tsing Hua University, Hsinchu 300044, Taiwan}

\author{Yen-Hsiang Lin}
\affiliation{Department of Physics, National Tsing Hua University, Hsinchu 300044, Taiwan}
\affiliation{Center for Quantum Science and Technology, National Tsing Hua University, Hsinchu 300044, Taiwan}
\affiliation{National Center for Excellence in Quantum Information Science and Engineering, National Tsing Hua University, Hsinchu 300044, Taiwan}

\author{Yung-Fu Chen}
\email[Correspondence to: ]{yfuchen@ncu.edu.tw}
\affiliation{Department of Physics, National Central University, Taoyuan 320317, Taiwan}
\affiliation{National Center for Excellence in Quantum Information Science and Engineering, National Tsing Hua University, Hsinchu 300044, Taiwan}
\affiliation{Center of High Energy and High Field Physics, National Central University, Taoyuan 320317, Taiwan}
\affiliation{Quantum Technology Center, National Central University, Taoyuan 320317, Taiwan}
\affiliation{Taiwan Semiconductor Research Institute, National Institutes of Applied Research, Hsinchu 300091, Taiwan}
\affiliation{Research Center for Critical Issues, Academia Sinica, Tainan 711010, Taiwan}

\date{\today}

\begin{abstract}
We theoretically investigate quantum-light storage in an effective $\Lambda$-type system formed by a pair of tunable superconducting qubits coupled to a semi-infinite transmission line and separated by half a wavelength. The frequency detuning between the qubits directly couples the superradiant excited state and the subradiant metastable state. By adjusting the qubit detuning, the system exhibits a controllable spectral response accompanied by slow-light behavior. Dynamically controlling the detuning enables an incident quantum field to be mapped onto the subradiant collective state, stored as a long-lived excitation, and subsequently retrieved on demand. Our results establish the proposed system as a minimal and tunable quantum-memory architecture for scalable superconducting quantum networks.
\end{abstract}
\maketitle

Superconducting quantum circuits provide a promising platform for quantum information processing~\cite{gu2017microwave,krantz2019quantum,blais2021circuit,arute2019quantum,abughanem2025superconducting,awschalom2025challenges} and distributed quantum networks~\cite{kimble2008quantum,wehner2018quantum,cacciapuoti2019quantum,caleffi2024distributed}. In such networks, microwave photons carry quantum information between spatially separated nodes, motivating coherent interfaces between flying photons and stationary quantum systems. Advances in waveguide quantum electrodynamics have enabled deterministic state transfer and remote entanglement using shaped microwave photons~\cite{kurpiers2018deterministic,axline2018demand,magnard2020microwave}. Extending these capabilities to the reversible storage and on-demand retrieval of propagating quantum states provides a quantum memory, an essential resource for synchronizing and distributing quantum information across a network~\cite{lvovsky2009optical,heshami2016quantum}.

Typically, quantum memories map propagating photons onto long-lived stationary excitations. In superconducting circuits, two main approaches have been demonstrated. The first employs an array of spectrally separated resonators, with photon absorption followed by atomic-frequency-comb-like rephasing~\cite{matanin2023toward,bao2021demand,makihara2024parametrically,matanin2026superconducting}. Efficient absorption requires impedance matching, constraining the resonator frequencies and couplings. Storage and retrieval are controlled either by aligning and subsequently redistributing the resonator frequencies to halt and resume relative phase evolution, or by switching the array’s coupling to the external transmission line. The second approach maps an incident photon onto a metastable state of an artificial $\Lambda$-type system using a strong coherent control field~\cite{chu2025slow,chu2023three,chen2026slowing,chu2026Highly}. Efficient storage requires two-photon resonance, with storage and retrieval controlled by switching the control field off and on, respectively. The $\Lambda$-type approach offers a relatively direct storage mechanism by coherently mapping the photon onto a metastable state, while precise control of the strong coherent field remains a key practical requirement.

In this work, we investigate a quantum memory based on an effective $\Lambda$-type system that enables photon storage without an additional strong coherent microwave control field. This $\Lambda$-type configuration is formed by the superradiant and subradiant collective states of a detuned superconducting qubit pair coupled to a common transmission line and separated by half a wavelength ($\lambda/2$-DQP). The superradiant state couples efficiently to propagating photons, whereas the subradiant state suppresses radiative decay and thus serves as a long-lived storage state. The detuning between the qubits provides the coherent coupling between the two collective states~\cite{kockum2018decoherence,ask2020synthesizing,soro2022chiral,shah2024stabilizing,almanakly2026driven} and therefore plays the role of the control coupling in conventional $\Lambda$-type schemes. By dynamically controlling the detuning, a propagating photon can be mapped onto the subradiant state for storage and subsequently retrieved. For a probe resonant with the superradiant-state transition, the two-photon-resonance condition is preserved during dynamic detuning control, in contrast to the parametric-modulation-based $\Lambda$-type scheme~\cite{chu2025slow,chu2025coherent}, where maintaining the two-photon-resonance condition requires additional compensation for modulation-induced shifts of the effective transition frequencies. Moreover, the proposed scheme requires only two qubits rather than an array of spectrally separated resonators as in multiresonator memories~\cite{matanin2023toward,bao2021demand,makihara2024parametrically,matanin2026superconducting}. In the following, we first show how the $\lambda/2$-DQP can be described as an effective $\Lambda$-type system in the collective-state basis. We then demonstrate single-photon slow light under a fixed qubit detuning, arising from the strong dispersive response, followed by quantum-light storage and retrieval through dynamic control of the qubit detuning. The $\lambda/2$-DQP thus provides a compact quantum-memory architecture based on controllable collective-state dynamics.

\begin{figure}[t!]
    \centering
    \includegraphics[width=0.36\textwidth]{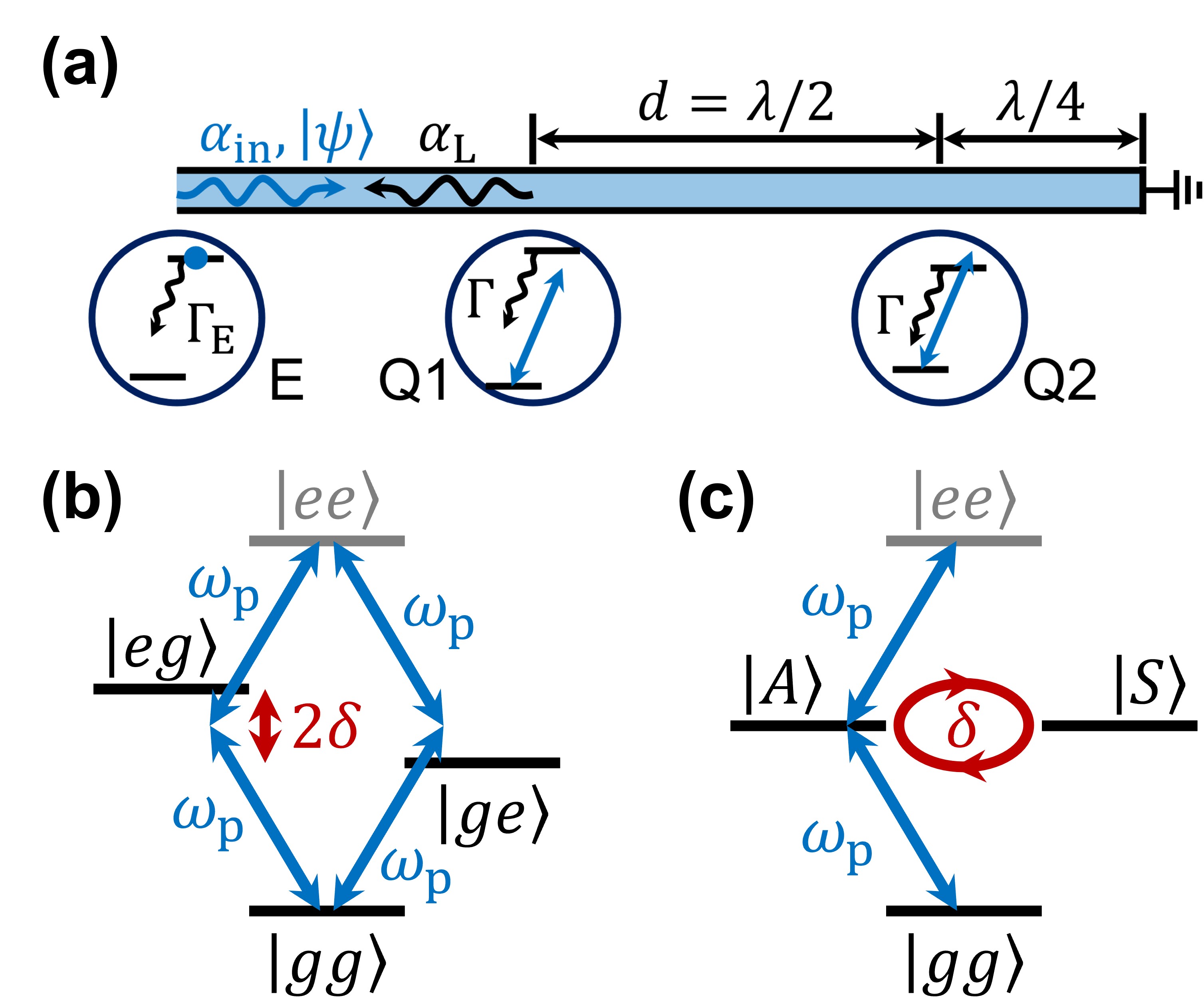}
    \caption{Effective $\Lambda$-type system formed by a $\lambda/2$-DQP. (a) Schematic of two qubits $Q_1$ and $Q_2$ coupled to a common semi-infinite transmission line, separated by $d=\lambda/2$, with $Q_2$ located $\lambda/4$ from the grounded termination. An ancillary emitter $E$ serves as the quantum-light source, while $\alpha_{\rm in}$ and $\alpha_{\rm L}$ denote the input and reflected fields, respectively. (b) Energy-level diagrams in the bare-qubit basis. The single-excitation states $|eg\rangle$ and $|ge\rangle$ are separated by $2\delta$, while the probe field at frequency $\omega_{\rm p}$ drives the transitions indicated by the blue arrows. (c) Corresponding collective-state basis. The probe couples the $gg-A$ transition, while $\delta$ coherently couples the $A-S$ transition.}
    \label{fig:DQP}
\end{figure}

Figure \ref{fig:DQP}(a) depicts two superconducting qubits, Q1 and Q2, coupled to a common semi-infinite transmission line. Each qubit has a ground state $|g\rangle$ and an excited state $|e\rangle$, with transition frequencies $\omega_{1,2}=\omega_0\pm\delta$, where $\omega_0$ is the average qubit frequency and $\delta$ is half the qubit frequency detuning. The qubits are separated by a distance $d=\lambda/2$, where $\lambda=2\pi v/\omega_0$ and $v$ is the microwave propagation velocity. The radiative decay rate of $Q_{\rm i}$ is modified by the standing-wave profile according to $\Gamma_{\rm i}=2\Gamma_{\rm b}\sin^2\left(\omega_{\rm i}L_{\rm i}/v\right)$, where $\Gamma_{\rm b}$ is the bare radiative decay rate and $L_{\rm i}$ is the distance from $Q_{\rm i}$ to the grounded termination~\cite{hoi2015probing}. The two qubits are placed at $L_1=3\lambda/4$ and $L_2=\lambda/4$, respectively, corresponding to antinodes of the standing-wave field. For $\delta\ll\omega_0$, the frequency dependence of the decay rate is negligible, yielding $\Gamma_1\approx\Gamma_2\equiv\Gamma=2\Gamma_{\rm b}$, while the residual mirror-induced Lamb shifts also remain negligible. Although direct coupling between the qubits is neglected due to their large separation, photon-mediated interactions through the common transmission line should be considered~\cite{van2013photon,lalumiere2013input}. For $d=\lambda/2$, the exchange coupling vanishes, while the correlated radiative decay rate is $\Gamma_{12}=\Gamma_{21}^*=-\Gamma e^{-i\delta d/v}$.

An input field $\alpha_{\rm in}$ is first treated as a coherent probe with angular frequency $\omega_{\rm p}$ and Rabi amplitude $\Omega_{\rm p}$. The propagation delay between the qubits is $\tau=d/v$. For $\Gamma\tau$, $\Omega_{\rm p}\tau\ll1$, the system operates in the Markovian regime~\cite{lalumiere2013input,gonzalez2011entanglement,zheng2013persistent}. Consequently, the propagation delay can be neglected, and the two qubits are regarded as being driven simultaneously by the incident field. The bare two-qubit states and probe-induced transitions are illustrated in Fig. \ref{fig:DQP}(b). In the frame rotating at $\omega_{\rm p}$ and under the rotating-wave approximation, the semiclassical interaction Hamiltonian is given by
\begin{equation}
\label{eq:Hint}
\begin{aligned}
    H &= \Delta_1\sigma_1^\dagger\sigma_1 + \Delta_2\sigma_2^\dagger\sigma_2\\
    &\quad-\frac{\Omega_{\rm p}}{2}\left( \sigma_1^\dagger+\sigma_1+e^{ik_{\rm p}d}\sigma_2^\dagger+e^{-ik_{\rm p}d}\sigma_2 \right),
\end{aligned}
\end{equation}
where $\sigma_i$ is the lowering operator of qubit $Q_i$, $\Delta_{\rm i}=\omega_{\rm i}-\omega_{\rm p}$, and $k_{\rm p}$ is the probe wavevector. The probe field acquires a relative phase $\pm k_{\rm p}d$ between the two qubits. The dissipative dynamics of the system are described by the Liouvillian
\begin{equation}
\label{eq:Liouvillian}
    \mathcal L(\rho) = \sum_{\rm i,j=1,2}\frac{\Gamma_{\rm ij}}{2}\mathcal D(\sigma_i,\sigma_j)\rho,
\end{equation}
where $\Gamma_{ii}=\Gamma$ is the individual radiative decay rate, $\rho$ is the system’s density matrix, and $\mathcal D(\sigma_{\rm i},\sigma_{\rm j})\rho=2\sigma_{\rm j}\rho\sigma_{\rm i}^\dagger-\left\{\sigma_{\rm i}^\dagger\sigma_j,\rho\right\}$ is the generalized Lindblad dissipator. Here, we first assume no qubit pure dephasing, $\Gamma_{\upphi}=0$, in deriving the effective $\Lambda$-type configuration.

To reveal the physical mechanism of the $\lambda/2$-DQP system, we employ the following approximations. First, for the correlated radiative decay rate, the additional propagation phase associated with the small qubit detuning $\delta\ll\omega_0$ is negligible, yielding $e^{\pm i\delta d/v}\approx1$. The correlated decay rates reduce to $\Gamma_{12}=\Gamma_{21}^*\approx-\Gamma$. The dissipative dynamics simplifies to a single collective decay channel, $\mathcal L(\rho)\approx \left(\Gamma/2\right)D\left( \sigma_1-\sigma_2 \right)\rho$, where $D\left( \sigma \right)\rho\equiv D\left( \sigma,\sigma \right)\rho$. Second, for the probe-field driving term in Eq. \ref{eq:Hint}, the propagation phase between the two qubits is approximately $\pi$ near $\omega_{\rm p}\approx\omega_0$, giving $e^{\pm ik_{\rm p}d}\approx-1$. We describe the system in the collective-state basis~\cite{dicke1954coherence,feng2017electromagnetically}, with the level structure shown in Fig. \ref{fig:DQP}(c). The antisymmetric and symmetric states are defined as
\begin{equation}
|A\rangle=\frac{|eg\rangle-|ge\rangle}{\sqrt{2}},\quad 
|S\rangle=\frac{|eg\rangle+|ge\rangle}{\sqrt{2}}.
\end{equation} 
In the weak-probe limit, $\Omega_{\rm p}\ll\Gamma$, the population of the doubly excited state $|ee\rangle$ is negligible, and the dynamics can be restricted to the three-state subspace spanned by $|gg\rangle$, $|A\rangle$, and $|S\rangle$. The Hamiltonian and Liouvillian then become
\begin{subequations}
\label{eq:3L}
\begin{align}
    \begin{split}
    H_{\rm 3L} &= \Delta_0\left( \sigma_{\rm A}^\dagger\sigma_{\rm A} + \sigma_{\rm S}^\dagger\sigma_{\rm S} \right)\\
    &\quad + \delta\left(\sigma_{\rm A}^\dagger\sigma_{\rm S} + \sigma_{\rm S}^\dagger\sigma_{\rm A}\right) - \frac{\Omega_{\rm p}}{\sqrt{2}}\left(\sigma_{\rm A}^\dagger + \sigma_{\rm A}\right),
    \end{split} \\
    \mathcal L(\rho)&=\frac{\Gamma_{\rm A}}{2}\mathcal D\left( \sigma_{\rm A} \right)\rho,
\end{align}
\end{subequations}
where $\Delta_0=\omega_0-\omega_{\rm p}$, $\sigma_{\rm A}=|gg\rangle\langle A|$ and $\sigma_{\rm S}=|gg\rangle\langle S|$ are the lowering operators associated with $|A\rangle$ and $|S\rangle$, respectively. The input field drives the $gg-A$ transition, while $\delta$ provides the coherent coupling between $|A\rangle$ and $|S\rangle$. The Liouvillian contains only the decay channel associated with $|A\rangle$, which forms a superradiant mode with decay rate $\Gamma_{\rm A}=2\Gamma$, whereas $|S\rangle$ forms a subradiant mode decoupled from the transmission line. The $\lambda/2$-DQP therefore realizes an effective $\Lambda$-type system. In this $\Lambda$-type description, $\Delta_0$ corresponds to both the single-photon and two-photon detunings. For a resonant field, $\omega_{\rm p}=\omega_0$, the two-photon detuning vanishes and remains unchanged as $\delta$ is varied.

The optical response of the $\lambda/2$-DQP system is expected to exhibit the characteristic spectrum of conventional $\Lambda$-type systems \cite{fleischhauer2005electromagnetically,sheremet2023waveguide}. To obtain the system dynamics, we numerically solve the Lindblad master equation using Eqs.~\ref{eq:Hint} and \ref{eq:Liouvillian}. Using the input-output formalism, the output field propagating to the left in Fig. \ref{fig:DQP}(a) is given by
\begin{equation}
\label{eq:InOut_end}
    \alpha_{\rm L} = \alpha_{\rm in} + i\sqrt{\Gamma}\left( \sigma_1-\sigma_2 \right),
\end{equation}
where $\alpha_{\rm in}=\Omega_{\rm p}/2\sqrt{\Gamma}$. Propagation phases between the qubits and those accumulated along the round-trip path involving reflection from the grounded termination are retained in the input-output relation~\cite{hoi2015probing}. For $d=\lambda/2$, $L_2=\lambda/4$, and $\omega_{\rm p}\approx\omega_0$, they reduce to a relative $\pi$ phase between the fields emitted by the two qubits. The reflection coefficient is defined as $R=\alpha_{\rm L}/\alpha_{\rm in}$, where the steady-state expectation values of the qubit operators are substituted into the input-output relation. The simulation parameters are $\omega_0/2\pi=5\ {\rm GHz}$, $\Gamma/2\pi=10\ {\rm MHz}$, and $\Omega_{\rm p}=0.01\Gamma$. The Markov approximation is satisfied as $\Gamma\tau=2\pi\times10^{-3}$ and $\Omega_{\rm p}\tau=2\pi\times10^{-5}$. 

\begin{figure}[t!]
    \centering
    \includegraphics[width=0.48\textwidth]{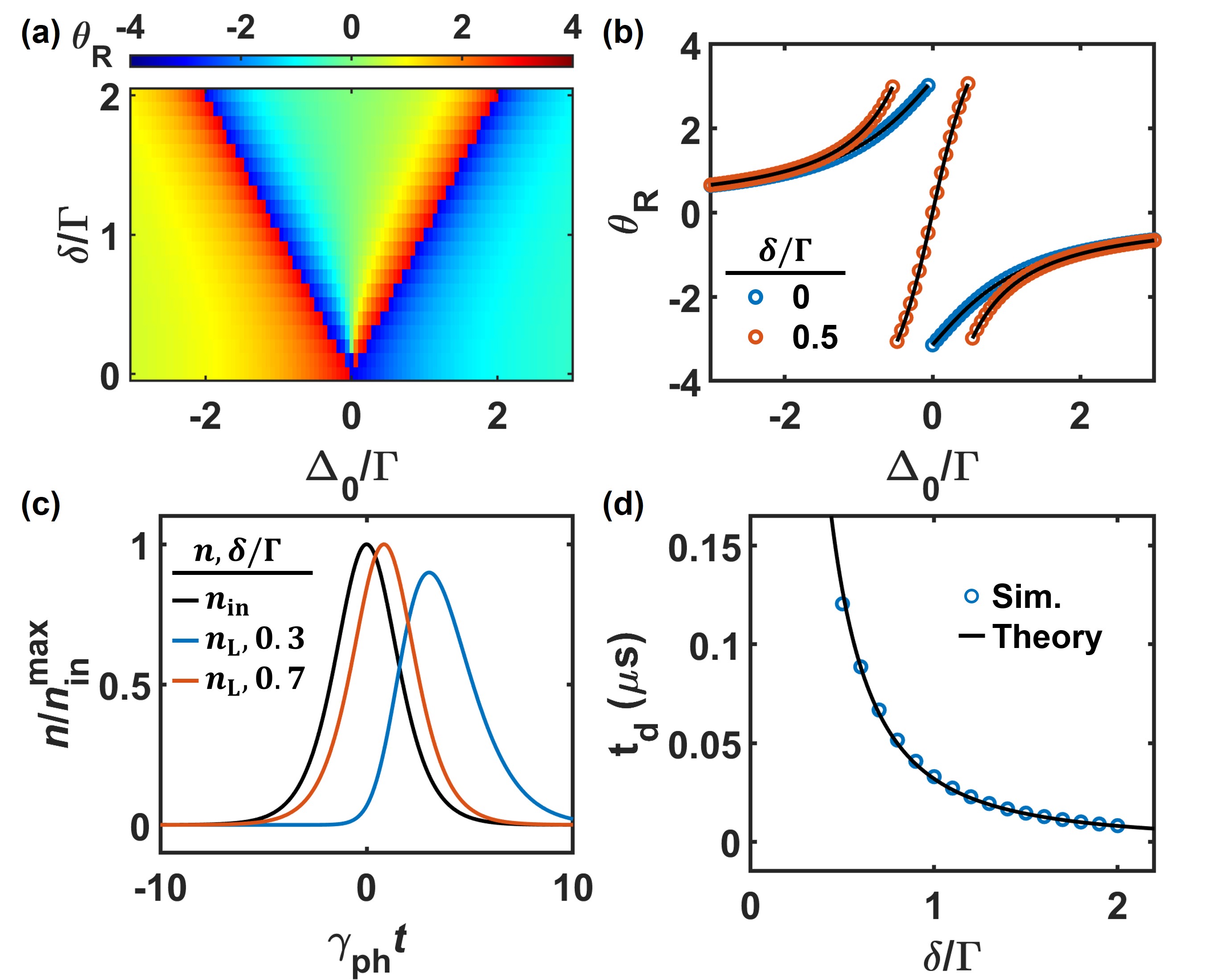}
    \caption{Optical response and single-photon slow light. (a) Reflection phase $\theta_{\rm R}$ as a function of $\Delta_0$ and $\delta$. (b) Line cuts of $\theta_{\rm R}$ at $\delta=0$ and $0.5\Gamma$. Symbols denote numerical simulations, while solid curves are the analytical results from Eq. \ref{eq:SIT}. (c) Input and reflected single-photon fluxes for $\gamma_{\rm ph}=0.2\Gamma$, demonstrating the slow-light effect. (d) Delay time $t_{\rm d}$ as a function of $\delta$. The theoretical result is given by Eq. \ref{eq:td}}
    \label{fig:SIT}
\end{figure}

The simulated reflection phase $\theta_{\rm R}=\arg(R)$ as a function of $\Delta_0$ and $\delta$ is shown in Fig. \ref{fig:SIT}(a). The evolution of $\theta_{\rm R}$ reflects the characteristic response of the effective $\Lambda$-type system, including interference- and splitting-dominated regimes~\cite{anisimov2011objectively,liu2016method,liu2016experimental}. A quantitative distinction between these regimes is provided in the Supplemental Material (SM), Sec. \textrm{I}~\cite{SM}. Figure \ref{fig:SIT}(b) shows $\theta_{\rm R}$ as a function of $\Delta_0$ for $\delta=0$ and $0.5\Gamma$. When the two qubits are resonant, $\delta=0$, the reflection phase exhibits a single resonant feature centered at $\Delta_0=0$. For $\delta=0.5\Gamma$, the reflection phase exhibits a steep dispersion around $\Delta_0=0$. In the collective-state basis, the output field is determined solely by the superradiant mode, $\alpha_{\rm L}=\alpha_{\rm in}+i\sqrt{\Gamma_{\rm A}}\sigma_{\rm A}$. Solving the steady-state Lindblad master equation associated with Eq. \ref{eq:3L} and substituting the resulting expectation value of $\sigma_{\rm A}$ into the input-output relation yields the analytical reflection coefficient
\begin{equation}
\label{eq:SIT}
    R = \frac{\left(\Delta_0+\delta\right)\left(\Delta_0-\delta\right)+i\Gamma\Delta_0}{\left(\Delta_0+\delta\right)\left(\Delta_0-\delta\right)-i\Gamma\Delta_0},
\end{equation}
which has the same steady-state solution as a $\Lambda$-type system \cite{fleischhauer2005electromagnetically,sheremet2023waveguide}. The excellent agreement between the analytical results and the numerical simulations in Fig. \ref{fig:SIT}(b) confirms that the effective $\Lambda$-type model captures the essential physics of the reflection response.

Beyond the semiclassical spectral response, the effective $\Lambda$-type configuration enables coherent manipulation of propagating quantum light, providing the basis for slow light and quantum-light storage. To describe these processes at the single-photon level, the system is extended to a full-quantum model of the qubits and propagating field. An ancillary qubit $E$ in Fig. \ref{fig:DQP}(a) serves as a deterministic quantum light emitter. The emitter is coupled only to the right-propagating mode, ensuring that it is not affected by the field reflected from the $\lambda/2$-DQP~\cite{gough2015generating,kiilerich2019input,gheeraert2020programmable}. The composite system consisting of the emitter and the $\lambda/2$-DQP is described by the Hamiltonian and Liouvillian
\begin{subequations}
\begin{align}
    \begin{split}
    H &= \Delta_1\sigma_1^\dagger\sigma_1 + \Delta_2\sigma_2^\dagger\sigma_2\\
    &\quad+ \frac{i}{2}\sqrt{\Gamma\Gamma_{\rm E}}\left( \sigma_{\rm E}^\dagger\sigma_1 - \sigma_1^\dagger\sigma_{\rm E}\right.\\
    &\quad\left.+ e^{-ik_{\rm p}d}\sigma_{\rm E}^\dagger\sigma_2-e^{ik_{\rm p}d}\sigma_2^\dagger\sigma_{\rm E} \right),
    \end{split} \\
    \begin{split}
    \mathcal L(\rho) &= \sum_{\rm i,j=1,2}\frac{\Gamma_{\rm ij}}{2}\mathcal D(\sigma_{\rm i},\sigma_{\rm j})\rho + \frac{\Gamma_{\rm E}}{2}\mathcal D(\sigma_{\rm E})\rho\\
    &\quad+ \frac{\sqrt{\Gamma\Gamma_{\rm E}}}{2}\left[ D(\sigma_{\rm E},\sigma_{\rm 1})\rho + D(\sigma_{\rm 1},\sigma_{\rm E})\rho\right.\\
    &\quad\left.+ e^{-ik_{\rm p}d}D(\sigma_{\rm E},\sigma_{\rm 2})\rho + e^{ik_{\rm p}d}D(\sigma_{\rm 2},\sigma_{\rm E})\rho \right],
    \end{split}
\end{align}
\end{subequations}
where $\Gamma_{\rm E}$ and $\sigma_{\rm E}$ are the radiative decay rate and lowering operator of the emitter, respectively. A shaped single-photon wave packet $u(t)=\frac{\sqrt{\gamma_{\rm ph}}}{2}{\rm sech}\left(\frac{\gamma_{\rm ph}t}{2}\right)$ is generated by controlling the time-dependent decay rate $\Gamma_{\rm E}(t)$, where $\gamma_{\rm ph}$ is the pulse bandwidth. See SM, Sec. \textrm{II}~\cite{SM} for details. The left-propagating output field $\alpha_{\rm L}$ and the corresponding photon flux $n_{\rm L}$ are obtained from the full-quantum input-output relation,
\begin{subequations}
\begin{align}
    \alpha_{\rm L} &= i\sqrt{\Gamma_{\rm E}}\sigma_{\rm E} + i\sqrt{\Gamma}\left( \sigma_1-\sigma_2 \right)\\
    n_{\rm L} &= \left\langle \alpha_{\rm L}^\dagger\alpha_{\rm L} \right\rangle.
\end{align}
\end{subequations} 
The input field and flux are given as $\alpha_{\rm in}=i\sqrt{\Gamma_{\rm E}}\sigma_{\rm E}$ and $n_{\rm in}=\Gamma_{\rm E}\sigma_{\rm E}^\dagger\sigma_{\rm E}$, respectively.

The steep reflection-phase dispersion near resonance in Fig. \ref{fig:SIT}(b) implies a large reflection group delay. The slow-light effect is demonstrated by an incident single-photon wave packet with $\gamma_{\rm ph}=0.2\Gamma$, starting with the emitter initialized in the excited state $|e\rangle_{\rm E}$ and the emitted photon resonant with $\omega_0$, i.e., $\Delta_0=0$. Figure \ref{fig:SIT}(c) shows the normalized photon flux $n_{\rm L}/n_{\rm in}^{\rm max}$ for different $\delta$, where $n_{\rm in}^{\rm max}$ is the maximum input photon flux. As expected from the phase response, a smaller $\delta$ results in a larger group delay due to the steeper phase dispersion around resonance. The frequency derivative of the reflection phase gives the group delay
\begin{equation}
\label{eq:td}
    t_{\rm d}=\left. \frac{\partial\theta_{\rm R}}{\partial\omega_{\rm p}}\right|_{\Delta_0=0}=2\frac{\Gamma}{\delta^2}.
\end{equation}
Figure \ref{fig:SIT}(d) summarizes the extracted group delay as a function of $\delta$. The analytical prediction of $t_{\rm d}$ agrees well with the numerical simulations. Comparison with the group-delay relation for a conventional $\Lambda$-type system~\cite{fleischhauer2005electromagnetically,sheremet2023waveguide}, $t_{\rm d}=D\frac{2\Gamma}{\left(2\delta\right)^2}$, yields an effective optical depth of $D=4$.

\begin{figure}[t!]
    \centering
    \includegraphics[width=0.48\textwidth]{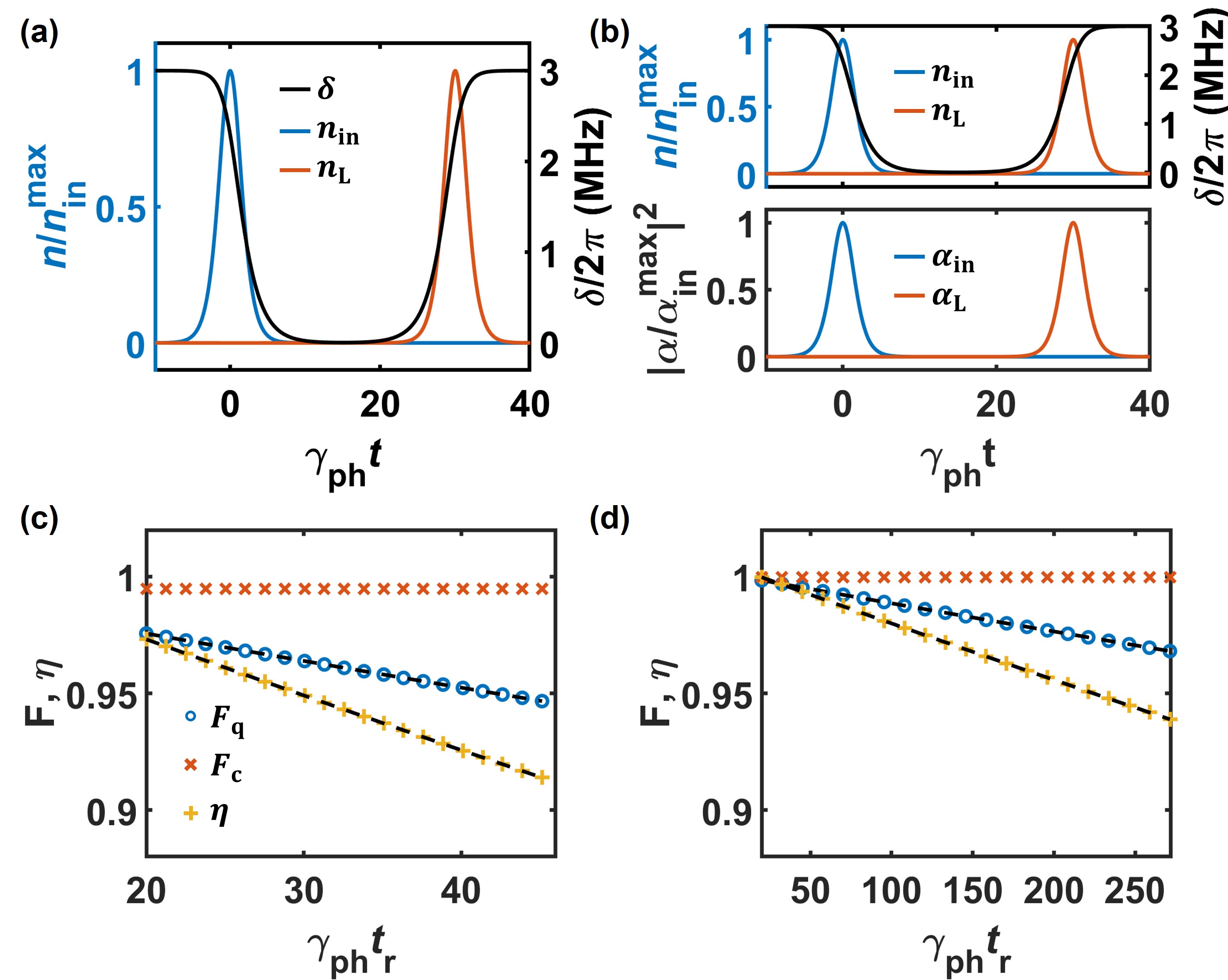}
    \caption{Quantum-light storage and retrieval. (a) Photon flux and detuning control for the storage and retrieval of a single-photon state. (b) Photon flux (top) and normalized squared field amplitudes (bottom) for the photonic superposition state. Panels (a) and (b) are obtained in the absence of dephasing. (c) Quantum-memory performance as a function of $t_{\rm r}$ under dephasing $\Gamma_\upphi/2\pi=5\ {\rm kHz}$ for $\gamma_{\rm ph}=0.2\Gamma$, and (d) for $\gamma_{\rm ph}=2\Gamma$. Black dashed lines show theoretical predictions based on exponential decay at rate $\Gamma_\upphi$.}
    \label{fig:QM}
\end{figure}

The effective $\Lambda$-type configuration enables quantum-light storage through dynamic control of the qubit detuning. Figure \ref{fig:QM}(a) illustrates the storage and retrieval of a single-photon flux with bandwidth $\gamma_{\rm ph}=0.2\Gamma$. Complete absorption of the target wave packet is achieved by dynamically controlling the qubit detuning as~\cite{kurpiers2018deterministic,SM,gheeraert2020programmable} 
\begin{equation}
\begin{aligned}
    \delta(t) &= \frac{\gamma_{\rm ph}}{4}{\rm sech}\left( \frac{-\gamma_{\rm ph}t}{2} \right) \\
    &\qquad\times\frac{1-e^{-\gamma_{\rm ph}t}+\left( 1+e^{-\gamma_{\rm ph}t} \right)\frac{2\Gamma}{\gamma_{\rm ph}}}{\sqrt{\left( 1+e^{-\gamma_{\rm ph}t} \right)\frac{2\Gamma}{\gamma_{\rm ph}}-e^{-\gamma_{\rm ph}t}}}.
\end{aligned}
\end{equation}
For $\gamma_{\rm ph}\leq\Gamma_{\rm A}=2\Gamma$, the photon is absorbed through $|A\rangle$ and coherently transferred to $|S\rangle$, suppressing radiative loss during storage. After a desired retrieval time $t_{\rm r}$, the time-reversed control $\delta(-t+t_{\rm r})$ transfers the excitation back to $|A\rangle$ and re-emits the photon into the transmission line. Figure \ref{fig:QM}(b) demonstrates the coherent storage for the photonic superposition state $|\psi\rangle=\left( |0\rangle+i|1\rangle \right)/\sqrt{2}$, generated by preparing the emitter in $\left( |g\rangle_{\rm E}+|e\rangle_{\rm E} \right)/\sqrt{2}$. The retrieved field reproduces both the photon flux and squared field amplitude of the incident wave packet, consistent with preserving photonic coherence during storage and retrieval.

The memory performance of the photonic superposition state is further evaluated in the presence of pure dephasing, which is particularly relevant for frequency-tunable superconducting qubits~\cite{braumuller2020characterizing,kumar2016origin}. The Liouvillian in Eq. \ref{eq:Liouvillian} includes pure dephasing through 
\begin{equation}
    \mathcal L_\upphi(\rho)=\Gamma_\upphi\left[ \mathcal D\left(\sigma_1^\dagger\sigma_1\right)\rho + \mathcal D\left(\sigma_2^\dagger\sigma_2\right)\rho \right].
\end{equation}
with $\Gamma_\upphi/2\pi = 5\ {\rm kHz}$ for each qubit, motivated by state-of-the-art superconducting-qubit coherence~\cite{almanakly2026driven}. The performance is characterized by the quantum fidelity $F_{\rm q}$, classical fidelity $F_{\rm c}$, and storage efficiency $\eta$, which quantify the preservation of the photonic quantum state, temporal waveform, and retrieved photon number, respectively. They are defined as 
\begin{subequations}
\begin{align}
    F_{\rm q}(\varrho_{\rm ret},\varrho_{\rm in})&=\left( {\rm Tr}\sqrt{\sqrt{\varrho_{\rm ret}}\varrho_{\rm in}\sqrt{\varrho_{\rm ret}}} \right)^2\\
    F_{\rm c}(n_{\rm L},n_{\rm in})&=\frac{\left|\int\sqrt{n_{\rm L}(t-t_{\rm r})n_{\rm in}(t)}dt\right|^2}{\int n_{\rm L}(t-t_{\rm r})dt\int n_{\rm in}(t)dt}\\
    \eta(n_{\rm L},n_{\rm in})&=\frac{\int n_{\rm L}(t-t_{\rm r})dt}{\int n_{\rm in}(t)dt},
\end{align}
\end{subequations}
where $\varrho_{\rm ret}$ is the density matrix of the retrieved photon and $\varrho_{\rm in}=|\psi\rangle\langle\psi|$ is that of the incident photon with $|\psi\rangle=\left( |0\rangle+i|1\rangle \right)/\sqrt{2}$. Detailed analysis is given in the SM, Sec. \textrm{III}~\cite{SM}. Figure \ref{fig:QM}(c) and (d) show these quantities of the retrieved photon as a function of $t_{\rm r}$ for $\gamma_{\rm ph}=0.2\Gamma$ and $2\Gamma$, respectively. Dephasing reduces the relative coherence between $|eg\rangle$ and $|ge\rangle$, inducing population mixing between $|S\rangle$ and $|A\rangle$. The stored excitation leaks through the superradiant state even when $\delta=0$. This leakage, together with the loss of coherence between $|gg\rangle$ and $|S\rangle$, reduces $F_{\rm q}$ and $\eta$  with increasing $t_{\rm r}$. Their decay is well described by the theoretical curves based on exponential decay at rate $\Gamma_\upphi$. For each $\gamma_{\rm ph}$, $F_{\rm c}$ remains nearly independent of $t_{\rm r}$, but is slightly lower for $\gamma_{\rm ph}=0.2\Gamma$ because dephasing-induced mixing accumulates over the longer emission duration and perturbs the retrieved waveform. At the shortest $t_{\rm r}$, $\gamma_{\rm ph}=0.2\Gamma$ yields $F_{\rm q}\approx0.98$, $F_{\rm c}\approx0.99$, and $\eta\approx0.97$, whereas $\gamma_{\rm ph}=2\Gamma$ gives $F_{\rm q}\approx F_{\rm c}\approx\eta\approx1$, owing to the shorter capture and retrieval processes and reduced dephasing accumulation.

In conclusion, we have demonstrated a quantum memory based on an effective $\Lambda$-type system formed by a pair of detuned superconducting qubits coupled to a semi-infinite transmission line and separated by half a wavelength. The qubit detuning coherently couples the superradiant and subradiant collective states, providing a tunable control parameter for their interconversion. The resulting dispersive response enables single-photon slow light, while dynamic control of the qubit detuning allows propagating quantum states to be mapped onto the long-lived subradiant state and subsequently retrieved on demand. The effective $\Lambda$-type configuration further supports high-fidelity remote-entanglement preparation through dark-state manipulation via stimulated Raman adiabatic passage (STIRAP)~\cite{vitanov1999creation,kumar2016stimulated,bergmann1998coherent,vitanov2001laser}, as detailed in the SM, Sec. \textrm{IV}~\cite{SM}. These results establish the $\lambda/2$-DQP as a compact platform for quantum memory and coherent collective-state manipulation in superconducting quantum networks.

\begin{acknowledgments}
We are grateful to Dr. Ching-Ping Lee and Prof. Wen-Te Liao for their helpful discussions. This work was supported by the National Science and Technology Council (NSTC), Taiwan, through Grants No. NSTC 114-2112-M-008-022, NSTC 114-2811-M-008-015, and NSTC 115-2112-M-008-009, and by the Taiwan Centers of Excellence program of the Ministry of Education, Taiwan.
\end{acknowledgments}

\end{document}